\documentclass[prl,twocolumn,showpacs,preprintnumbers,amsmath,amssymb]{revtex4}
\usepackage{graphicx}% Include figure files
\usepackage{dcolumn}% Align table columns on decimal point
\usepackage{bm}% bold math
\usepackage[tight]{subfigure}
\usepackage{enumerate}
\usepackage{amsmath}
\usepackage{verbatim}
\usepackage{color}

\begin{document}
%Only a first suggestion for a title
\title{Thermoelectric characterization of the Kondo resonance in nanowire quantum dots}
\author{Artis Svilans}
\author{Martin Josefsson}
\author{Adam M. Burke}
\author{Sofia Fahlvik}
\author{Claes Thelander}
\author{Heiner Linke}
\author{Martin Leijnse}

\affiliation{Division of Solid State Physics and NanoLund, Lund University, Box 118,S-221 00 Lund, Sweden}

\begin{abstract}

We experimentally verify hitherto untested theoretical predictions about the thermoelectric properties of Kondo correlated quantum dots (QDs). The specific conditions required for this study are obtained by using QDs epitaxially grown in nanowires, combined with a recently developed method for controlling and measuring temperature differences at the nanoscale. This makes it possible to obtain data of very high quality both below and above the Kondo temperature, and allows a quantitative comparison with theoretical predictions. Specifically, we verify that Kondo correlations can induce a polarity change of the thermoelectric current, which can be reversed either by increasing the temperature or by applying a magnetic field.

\end{abstract}
\pacs{
72.10.Fk, % Kondo
72.20.Pa, %  Thermoelectric
73.63.Kv, % QD transport
81.07.Gf % Nanowires
%84.60.Rb, % Thermoelectric energy conversion
%62.23.Hj, % Nanowires
%73.63.Nm % Quantum wires, electronic transport in
}

\maketitle

Measurements of electric and thermoelectric transport properties can be used to reveal and characterize novel strongly correlated phases, which often appear in meso- and nano-scale systems. The Kondo effect \cite{Kondo64} is a prominent example where interactions between conduction electrons and magnetic impurities result in a many-body singlet state involving the impurity spin and a large number of conduction electrons. In metals it leads to increased resistivity at low temperatures where the magnetic impurity scattering dominates. 
More recently, quantum dots (QDs) tunnel-coupled to two leads have provided a platform for more detailed experimental studies of the Kondo effect~%\cite{Goldhaber98,Cronenwett98,Schmid98,Simmel99,Scheibner05,Li98,Park02,Liang02,Yu04c60,YuKeane05,Scott09,Nygard00,Quay07,Jespersen06,Csonka08,Nilsson09,Kretinin10,Kretinin11,Kretinin12,Das12,Finck13,Scott13,Hemingway14,Petit14,Kanai17}.
\cite{Goldhaber98,Cronenwett98,Nygard00}. In QDs, the Kondo scattering lifts the Coulomb blockade~\cite{Ng88,Glazman88,Hewson93} and gives rise to a peak in the differential conductance $g=dI/dV$ around $V=0$ ($I$ is the current and $V$ is the bias voltage).

Several theoretical works (see, e.g., Refs~\cite{Boese01b,Costi10,RouraBas2012,Azema2012,Weymann2013,Ye14,Dorda2016,Sierra17,Karki2017}) have proposed that additional insights into Kondo physics can be gained from thermoelectric measurements. Here, a temperature difference $\Delta T = T_c - T_h$ is applied between a hot ($h$) and a cold ($c$) lead and one measures either the resulting thermocurrent $I_{th}$ (measured under closed-circuit conditions), or the thermovoltage $V_{th}$ (measured under open-circuit conditions). In QDs without Kondo correlations, $I_{th}$ and $V_{th}$ have characteristic shapes as functions of the gate voltage, $V_G$, exhibiting a sign reversal (zero crossing) at each charge degeneracy point as well as in the center of each Coulomb valley~\cite{Beenakker92, Staring93, Dzurak93}. It has been theoretically predicted~\cite{Costi10} that Kondo correlations would significantly change this behavior by removing some zero crossings and consequently reversing the polarity of $I_{th}$ and $V_{th}$ over a finite $V_G$ range. Whether a QD shows the typical Kondo or non-Kondo behavior depends sensitively on several system parameters. Therefore, by observing the qualitative change in thermoelectric response as Kondo correlations are suppressed, e.g., by increased average temperature $T=(T_h+T_c)/2$ or magnetic field $B$, one can not only gain insights into Kondo physics, but also probe the internal QD energy scales. 

Despite such clear theoretical predictions, experimental studies of the thermoelectric properties of Kondo correlated QDs remain rather limited~\cite{Scheibner05,Svensson13}. Experimentally uncontrolled internal QD degrees of freedom often complicate even a qualitative comparison with theory. Therefore, the predicted reversal of $I_{th}$ and $V_{th}$ has been difficult to observe (although some unpublished data exist~\cite{ThierschmannPhDth}) and the response to a $B$ field has, to the best of our knowledge, not been investigated. 

In this Letter, we take important steps towards filling this gap between experiments and theory by presenting thermoelectric measurements on several Kondo correlated QDs. We measure $I_{th}$ and $g(V \approx 0)=g_0$ over consecutive Kondo and non-Kondo Coulomb valleys. We observe the sign reversal of $I_{th}$ theoretically predicted for the Kondo regime, and measure the transition between Kondo and non-Kondo behavior as $T$ is increased, finding quantitative agreement with theory~\cite{Costi10}. Furthermore, we apply an external $B$ field which also destroys Kondo correlations and find that a surprisingly large $B$ is needed to recover the typical non-Kondo behavior.

Our observations necessitate overcoming significant experimental challenges to access the parameter regimes which most clearly reveal the Kondo correlations and allow detailed comparison with theory predictions. The requirements include:
{\it (i)} strong quantum confinement such that transport is dominated by a single orbital ($\delta \epsilon \gg k_B T, \Gamma$, where $\delta \epsilon$ is the orbital spacing and $\Gamma$ is the tunnel coupling);
{\it (ii)} large charging energy $U$ and tunnel coupling, and low temperature, such that the Kondo regime is reached ($U \gg \Gamma \gg k_B T$ and $T < T_K$, where $T_K$ is the Kondo temperature);
{\it (iii)} application and characterization of $\Delta T$ across a very small QD;
{\it (iv)} the ability to controllably tune the system in and out of the Kondo regime, e.g., by gating or by varying $T$ or $B$.

\begin{figure}\hspace{-6 mm}
 \centering
 \includegraphics[width=1.05\linewidth]{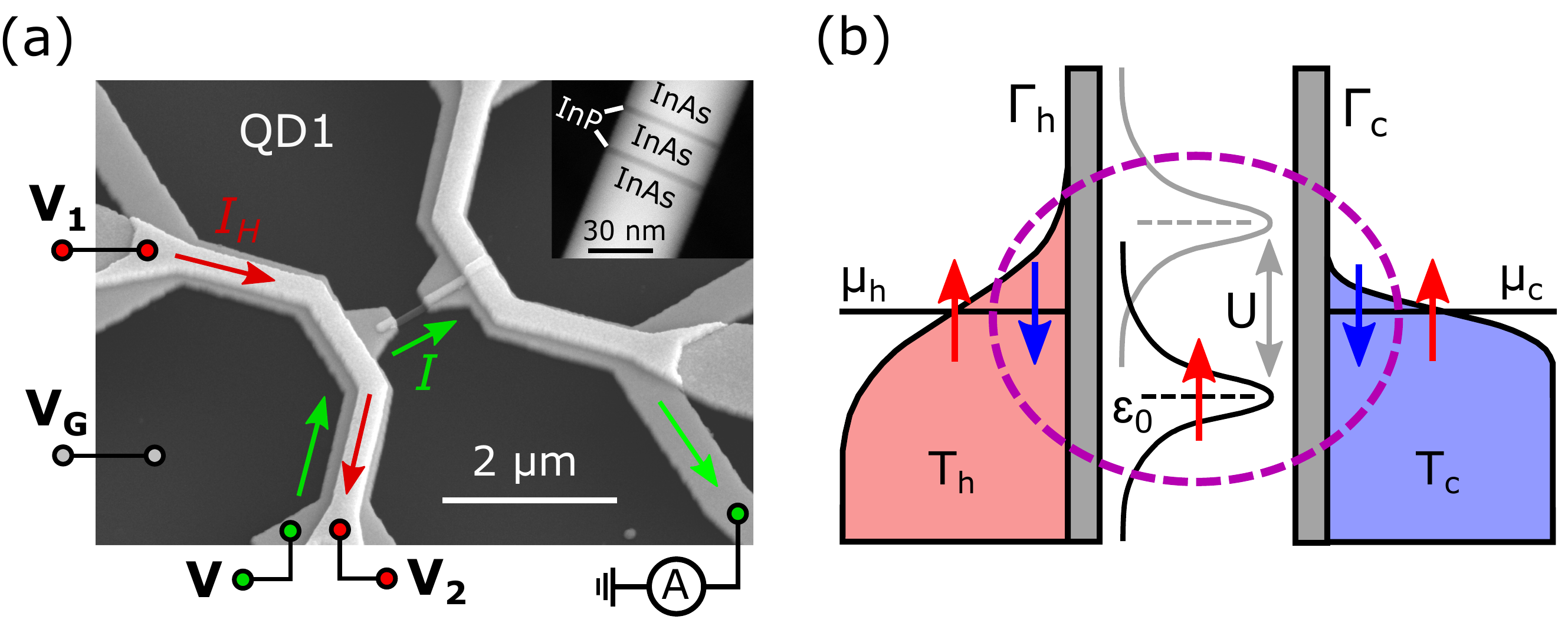}	
 \caption{(a) 
Scanning electron microscope (SEM) image of device QD3. An InAs/InP nanowire containing a QD is contacted to metallic leads for electrical biasing with voltage $V$ (see SM~\cite{SM} for details on circuitry). Additional heater leads (lighter gray) enable application of a thermal bias $\Delta T$ to the QD by running a current $I_H$ resulting from a heater bias $V_H = V_1 - V_2$. Only one heater is used in the experiment. Inset: Close-up scanning transmission electron microscope with high angle annular dark field (STEM-HAADF) image of an InAs/InP nanowire from the same growth. (b) Sketch of an unbiased spin-1/2 QD tunnel-coupled to two leads ($h$ and $c$).
}
 \label{fig:Fig1}
\end{figure}

To achieve the above requirements we use QDs epitaxially defined in axially heterostructured InAs/InP nanowires grown by chemical beam epitaxy \cite{Bjork04} [see inset in Fig.~\ref{fig:Fig1}(a)]. Each InAs nanowire from the same growth is about 60 nm in diameter and contains two thin InP segments that confine an approximately 20 nm long InAs QD, similar to those used in our previous studies \cite{Svilans16,Josefsson&Svilans17}. The small QD size and the small effective mass of InAs give sufficiently large $U$ and $\delta \epsilon$, and the large g-factor allows tuning the Zeeman energy over a wide range. We use the fabrication process developed in Ref.~\cite{Gluschke14} to fabricate thermoelectric devices. Figure~\ref{fig:Fig1}(a) shows a scanning electron microscope (SEM) image of the device QD1. In short, the devices are fabricated on an \textit{n}-doped Si wafer coated with $\mbox{SiO}_2$. Two Ni/Au leads are used to contact the outer InAs segments on each side of the QD. The nanowires along with the contacting leads are coated with $\mbox{HfO}_2$ in order to electrically isolate the heater leads from the electrical biasing circuit \cite{Gluschke14,Svilans16}. A back contact to the Si wafer is at a voltage $V_G$ and allows for electrostatic gating of the epitaxially defined QDs. We let $T_h$ and $T_c$ denote the temperatures of the nanowire leads contacting the QD, which might differ from those in the metallic leads further away. Application of a heating current $I_H$ increases both $T_h$ and (to a lesser degree) $T_c$, which in our devices gives control over $\Delta T$ and $T$ while maintaining a roughly constant $\Delta T/T\approx 0.30 - 0.35$. $T_h$ and $T_c$ are estimated based on QD thermometry (see Supplemental Material (SM)~\cite{SM} and Ref.~\cite{Josefsson&Svilans17} for details). During this study we characterized three QDs (QD1, QD2 and QD3) showing similar behavior. Only the data from QD1 is presented with figures in the main text. Results on all devices are summarized in Table \ref{Tab} (see SM~\cite{SM} for the corresponding data on other devices). The characterization was done in a dilution refrigerator with electron base temperature $T_0 < 100 \mbox{ mK}$. 

Figure \ref{fig:Fig1}(b) shows a sketch of a single-level QD with orbital energy $\varepsilon_0$ and onsite Coulomb repulsion $U$, coupled to leads by tunnel couplings $\Gamma_h$ and $\Gamma_c$ (Anderson model). The Kondo effect occurs when the level is occupied by a single electron. It originates from anti-ferromagnetic exchange interaction due to virtual exchange of electrons between the leads and the QD. Kondo correlations give rise to the formation of a singlet-like state (with binding energy $\sim k_B T_K$), involving the QD spin and a large number of electron spins in the leads. Below this energy the system behaves as a Fermi liquid and Coulomb blockade is lifted. 
\begin{figure}\hspace{-4 mm}
 \centering
 \includegraphics[width=0.93\linewidth]{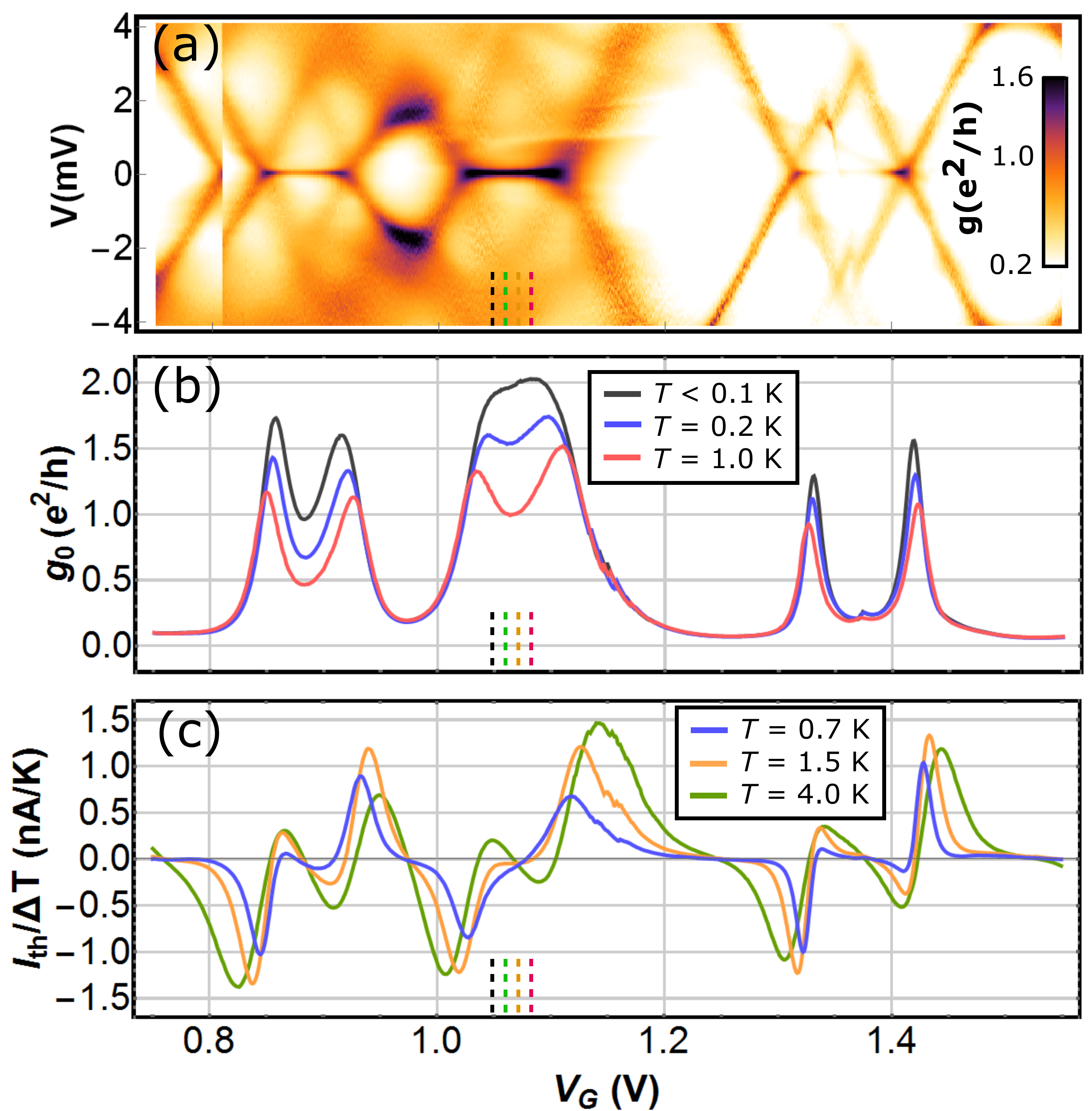}	
 \caption{
(a) $g$ as a function of $V$ and $V_G$, measured at $T=T_0 <100$ mK. (b) The corresponding $g_0$ as a function of $V_G$ measured at three different $T$. (c) $I_{th}$ normalized by $\Delta T$ as a function of $V_G$ measured at three different $T$. The horizontal $V_G$ axis is the same in (a)--(c). Vertical dashed lines refer to $V_G$ values in Fig.~\ref{fig:Fig3}. 
}
 \label{fig:Fig2}
\end{figure}

We use Fig.~\ref{fig:Fig2} to identify the effects of Kondo correlations in the experimental data. The measured charge stability diagram at $T_0 < 100$ mK in Fig.~\ref{fig:Fig2}(a) shows an increased $g_0$ inside Coulomb diamonds corresponding to odd electron numbers on the QD. In the absence of Kondo correlations one expects $g_0 < e^2/h$, but Fig.~\ref{fig:Fig2}(b) shows that (at $T=T_0$) $g_0$ approaches the limit $2e^2/h$, as expected in the Kondo regime. Increasing $T$ reduces $g_0$ in the odd occupancy Coulomb valleys but has little effect on valleys with even occupancy. 

Figure~\ref{fig:Fig2}(c) shows $I_{th}/\Delta T$ measured over the same gate range. We note that around $V_G=1.06$~V, where the strongest Kondo correlations are seen in (a) and (b), there is a qualitative change in $I_{th}(V_G)$ with increasing $T$ with two sign reversals being absent at low $T$. Therefore, we focus our analysis on this particular $V_G$ range and come back to a detailed discussion of the thermoelectric behavior later.

\begin{figure}\hspace{-2 mm}
 \centering
 \includegraphics[width=0.93\linewidth]{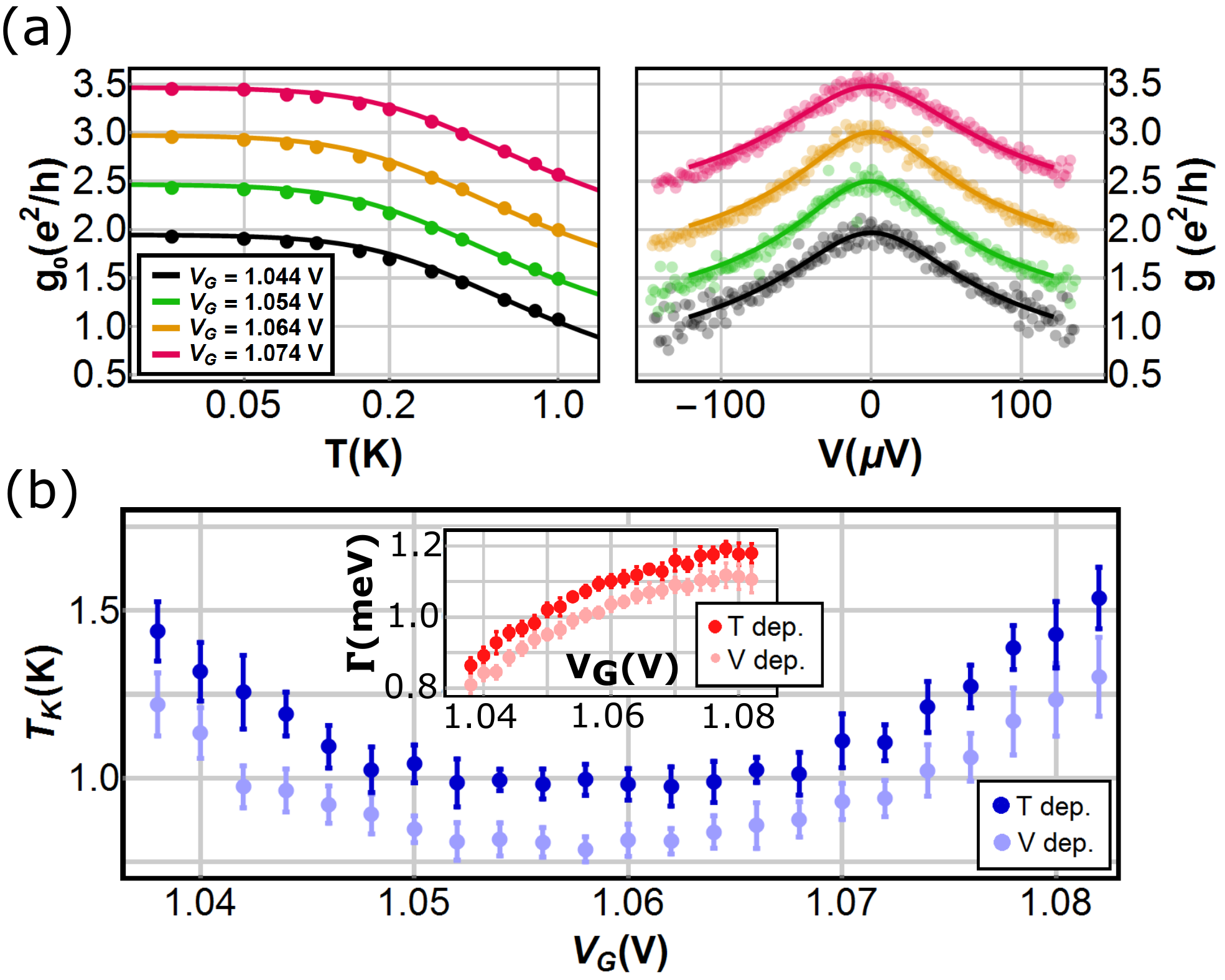}	
 \caption{(a) Dots are the measured values of $g_0(T)$ in the left panel and of $g(V)$ at $T_0$ in the right panel, both for four different values of $V_G$, also indicated by vertical dashed lines in Fig.~\ref{fig:Fig2} with the corresponding colors. The solid lines for $g_0(T)$ are fits to Eq.~(\ref{eq:g(T)}) whereas the solid lines for $g(V)$ are fits to Eq.~(2) in the SM~\cite{SM}. The curves and data points at different $V_G$ are offset by 0.5 $e^2/h$ in $g$ from each other. (b) $T_K$ as a function of $V_G$ determined from fitting the measured $g_0(T)$ to  Eq.~(\ref{eq:g(T)}) (dark blue points) and the measured $g(V)$ to Eq.~(2) in the SM~\cite{SM} (light blue points). The error bars represents a 95 \% confidence interval for $T_K$ as a fit parameter. Inset to (b): the corresponding estimates of $\Gamma$ as a function of $V_G$ using Eq.~(\ref{eq:TK}).}
 \label{fig:Fig3}
\end{figure}

Figure~\ref{fig:Fig3} presents the analysis for determining $T_K$ and $\Gamma$. We use the $T$ dependence of $g_0$ at the chosen $V_G$ range to determine $T_K$ using the phenomenological expression~\cite{Goldhaber98,Kretinin11}
\begin{equation}\label{eq:g(T)}
g_0(T) = g_0(T=0) \left[ 1 + \left(2^{1/s}-1\right)\left(\frac{T}{T_K}\right)^2\right]^{-s},
\end{equation}
where $s=0.22$ for a spin-$1/2$ QD and $g_0(T=0)$ and $T_K$ are used as free fit parameters. Examples of the fits can be seen in the left panel of Fig.~\ref{fig:Fig3}(a) while the corresponding $T_K$ fit values are plotted in Fig.~\ref{fig:Fig3}(b). We also cross check the $T_K$ values by fitting the $V$ dependence of $g$ instead~\cite{Pletyukhov2012, SM}. Examples of those fits are shown in the right panel of Fig.~\ref{fig:Fig3}(a) while the corresponding fit values are plotted in Fig.~\ref{fig:Fig3}(b). Overall, the two methods agree well, although the $V$ dependence of $g$ yields somewhat lower $T_K$ values.

For a single-orbital model, $T_K$ is given by~\cite{Hewson93}
\begin{equation}\label{eq:TK}
k_B T_K = \frac{1}{2}\sqrt{\Gamma U}\exp{\left({\frac{\pi \varepsilon_0(\varepsilon_0+U)}{\Gamma U}}\right)},
\end{equation}
where $\varepsilon_0$ is the energy of the QD orbital relative to the Fermi level of the leads and varies from $0$ to $-U$ across the Coulomb valley. Equation~(\ref{eq:TK}) is strictly valid only in the Kondo regime where $-U + \Gamma/2 < \varepsilon_0 < -\Gamma/2$ ~\cite{Haldane1978,Kretinin11}, i.e., far enough from the charge degeneracy points into the Coulomb valley. We estimate $U \approx 3.5$~meV which is used to calculate $\Gamma$ from the estimated $T_K$ values using Eq.~(\ref{eq:TK}) [see inset of Fig.~\ref{fig:Fig3}(c)]. We find that $\Gamma$ has a slight $V_G$ dependence, which is commonly observed in nanowire QDs because of the quasi one-dimensional density of states in the leads. 
\begin{figure*}\hspace{-4 mm}
 \centering
 \includegraphics[width=0.98\linewidth]{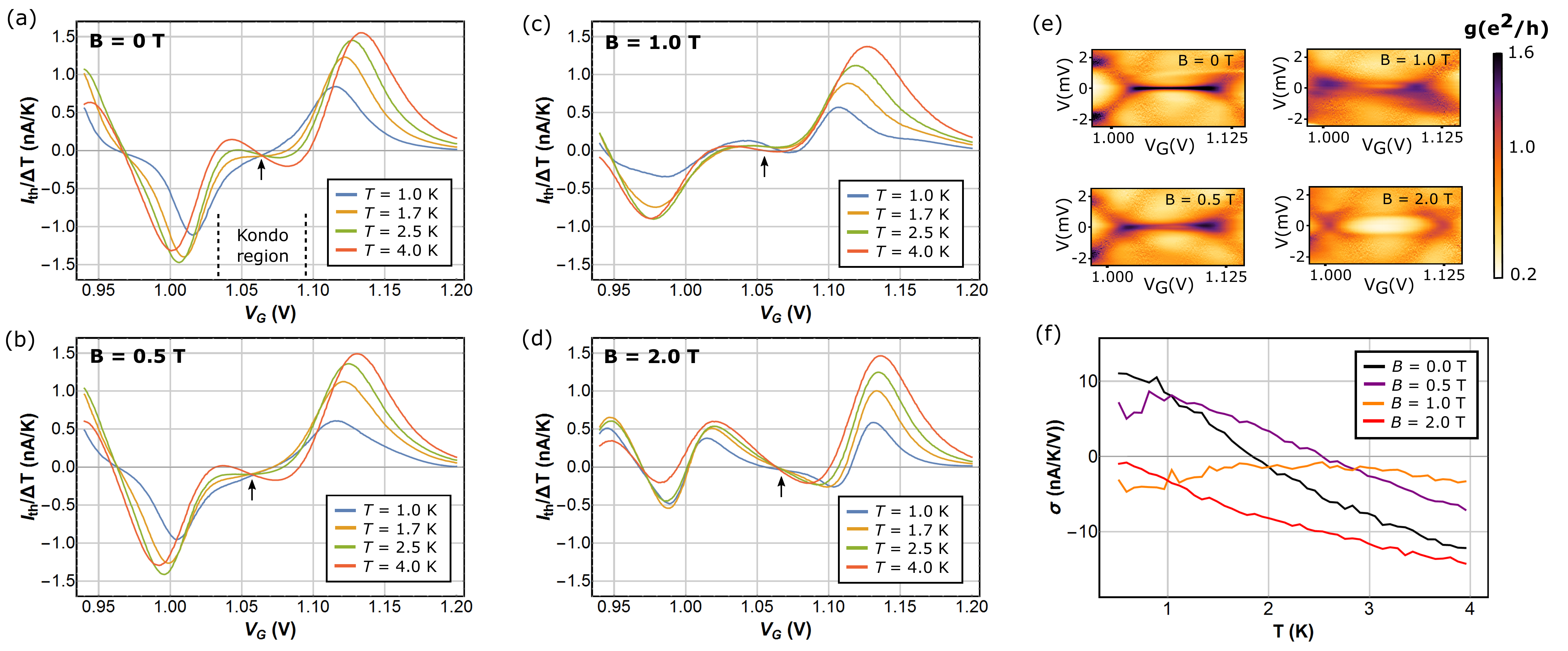}	
 \caption{(a)--(d) Measured $I_{th}$ normalized by $\Delta T$ for different $T$ as indicated in the figure. The black arrows indicate $V_G$ positions around which the values of $\sigma$ in (f) are determined. The magnetic field is increased from $B=0$ to $B=2$~T from (a)--(d) as indicated in the figures. The dashed lines in (a) indicate the $V_G$ range that corresponds to the Kondo regime. (e) Charge stability diagrams showing $g$ as a function of $V$ and  $V_G$, measured at the base temperature $T_0 <100$ mK for values of $B$ corresponding to those used in (a)--(d). (f) Thermocurrent slope, $\sigma = (dI_{th}/dV_G)/\Delta T$, as a function of $T$ at $B$ field values corresponding to those used in (a)--(d).}
 \label{fig:Fig4}
\end{figure*}

We now turn our attention to the thermoelectric properties of QDs in the Kondo regime. The $T = 4$~K trace in Fig.~\ref{fig:Fig2}(c) illustrates the expected behavior of $I_{th}(V_G)/\Delta T$ in the absence of Kondo correlations, where it undergoes twice as many sign reversals as there are charge degeneracy points -- one when passing through zero at each of the charge degeneracy points and one in the middle of every Coulomb valley~\cite{Beenakker92, Staring93, Dzurak93}. It was theoretically predicted in Ref.~\cite{Costi10} that this behavior is qualitatively different in the presence of Kondo correlations, which cause the zero crossings at the degeneracy points to disappear. Our experimental data verifies this prediction as the two sign reversals in the gate range between $V_G = 1.02$ and $1.10$~V disappear at low $T$. Two additional Kondo resonances are also seen in Fig.~\ref{fig:Fig2} (close to $V_G=0.9$ and $1.4$~V) but because $T_K$ is lower in those cases the sign reversal is not observed in Fig.~\ref{fig:Fig2}(c).

Figure~\ref{fig:Fig4} shows the sign reversal of $I_{th}$ more closely and focuses on the effects of increasing $T$ and $B$. Both are known to destroy Kondo correlations and it is therefore intuitive that also the sign reversal should be affected. We aim to quantify the values of $B$ and $T$ below which the Kondo-induced sign reversal takes place. Figures~\ref{fig:Fig4}(a)--(d) show data at different $B$ values, each for several different $T$. The corresponding charge stability diagrams for the same values of $B$ are displayed in Fig.~\ref{fig:Fig4}(e). The sign reversal of $I_{th}$ as a function of $T$ is best seen in Fig.~\ref{fig:Fig4}(a) where $B=0$~T. The trace at $T = 1~\mathrm{K} \approx T_K$ shows a single zero-crossing within the Coulomb valley marked as the "Kondo region". By raising the temperature $T > T_K$ the two additional zero-crossings are recovered, indicating a reversal of the direction of $I_{th}$ within the Kondo region. This observation is a clear verification of theoretical predictions in Ref.~\cite{Costi10}.

Based on the splitting of the Kondo peak, observable in Fig.~\ref{fig:Fig4}(e), we estimate the electron g-factor $|g_Z| \approx 9$~\cite{SM}. Thus, $T_K \approx 1$~K corresponds to $B \approx 0.17$~T. Interestingly, however, the behavior of $I_{th}$ at $B=0.5$~T, as shown in Fig.~\ref{fig:Fig4}(b), remains qualitatively and quantitatively similar to the zero field case. Only when increasing $B$ to 1.0~T and 2.0~T the two additional zero-crossings are recovered at all accessible $T$ [see Figs.~\ref{fig:Fig4}(c) and (d)].

Closer examination of the sign reversal requires analysis of the small $I_{th}$ within the Coulomb valley which is sensitive to the experimental uncertainties in the applied electrical bias ($\delta V\approx\pm 1$~$\mu$V). We therefore fit the gate-slope of the thermocurrent, $(dI_{th}/dV_G)/\Delta T = \sigma$, at the center of the Coulomb valley [marked by arrows in Figs.~\ref{fig:Fig4}(a)--(d)] and use its sign as an alternative indicator for the sign reversal of $I_{th}$. Figure~\ref{fig:Fig4}(f) shows $\sigma(T)$ measured at different $B$ values. We let $T_1$ denote the temperature at which $\sigma(T)$ changes from positive to negative. Interestingly, we find that $T_1$ is larger at $B=0.5$~T ($T_1 \approx 2.5$~K) than at $B = 0$~T ($T_1 \approx 1.8$~K), however this result does not seem to be reproduced in other devices and we do not have an explanation for it. In contrast, at field values $B=1.0$~T and $B=2.0$~T $\sigma$ no longer reverses sign as a function of $T$. Therefore, we conclude that the crossover happens between $B = 0.5$~T and $B = 1.0$~T. These field values correspond to $|g_Z| \mu_B B/k_B \approx 3 T_K - 6 T_K$, which is consistent with measurements under $B$ field on other devices \cite{SM}.

Table \ref{Tab} summarizes our results from all devices, see SM~\cite{SM} for the corresponding data and analysis. We estimate the relative errors for $U$ and $T_K$ to be in the range $\pm 10 \%$, which translates into a similar error for $\Gamma$. The accuracy of $T_1$ depends mostly on the accuracy of the thermometry, which we have not been able to quantify. However, we do not expect it to be a source of significant error. For all resonances we find $T_1/T_K \approx 1.2 - 1.8$. This is in good quantitative agreement with theory predictions in Ref.~\cite{Costi10} where $T_1/T_K \approx 1.6$ for $U/\Gamma = 3$.

\begin{table}
\caption{Summary of data from several devices. QD1a (in bold) represents results obtained on the device QD1 for which the data is shown in this Letter. QD1b represents results obtained on the same device QD1 but in a different $V_G$ range. QD2 and QD3 represents results obtained on devices QD2 and QD3.}\label{Tab}
\begin{ruledtabular}
\begin{tabular}{l*{6}{c}r}
                  &  $U$(meV)  \hspace{2mm}
                  &  $T_K$(K)  \hspace{2mm}
                  &  $\Gamma$(meV)  \hspace{2mm}
                  &  $T_1$(K)  & \\
%\hline \\
%\\
\textbf{QD1a}           & \textbf{3.5} & \textbf{1.0} & \textbf{1.1} & \textbf{1.8}  \\
QD1b          & 2.2 & 0.6 & 0.7 & 0.7  \\
QD2            & 2.6 & 0.6 & 0.8 & 0.8  \\
QD3           & 3.0 & 0.8 & 1.0 & 1.2  \\
\end{tabular}
\end{ruledtabular}
\end{table}

In conclusion, we have presented a detailed experimental study of the thermoelectric properties of Kondo correlated QDs. Our measurements confirm the theoretical prediction~\cite{Costi10} that sufficiently strong Kondo correlations can reverse the direction of $I_{th}$ over a finite range in $V_G$. We find quantitative agreement with theoretical predictions for the temperature $T_1$ at which sign the reversal takes place. We have also investigated the magnetic field dependence of $I_{th}$ and conclude that, unlike other transport quantities which change behavior at $|g_Z| \mu_B B / (k_B T_K ) \approx 1$ \cite{Filippone18}, the sign reversal of $I_{th}$ remains until this ratio is significantly larger than 1. This raises new questions and opens up for further theoretical and experimental studies. More generally, our work demonstrates that the use of thermoelectric measurements can be a sensitive probe of Kondo physics and other strong correlation effects. An interesting direction for future works is to investigate more complex QDs with additional symmetries~\cite{Karki2017} or the nonlinear, large $\Delta T$, regime~\cite{Sierra17, Dorda2016}, where theoretical predictions are much more challenging.

\acknowledgements
\emph{Acknowledgements} -- We gratefully acknowledge funding from the People Programme (Marie Curie Actions) of the European Union's Seventh Framework Programme (FP7-People-2013-ITN) under REA grant agreement no. 608153 (PhD4Energy), from the Swedish Research Council (projects 2012-5122 and 2016-03824), from the Knut and Alice Wallenberg Foundation (project 2016.0089), from the Swedish Energy Agency project (project 38331-1), from NanoLund, and computational resources from the Swedish National Infrastructure for Computing (SNIC) at LUNARC (projects SNIC 2017/4-10 and SNIC 2018/6-3).

\bibliography{cite}
\bibliographystyle{apsrev}
\end{document}

%Only a first suggestion for a title
\title{
Supplementary Material
\vspace{5mm}

Thermoelectric characterization of the Kondo resonance in nanowire quantum dots
}
\author{Artis Svilans}
\author{Martin Josefsson}
\author{Adam M. Burke}
\author{Sofia Fahlvik}
\author{Claes Thelander}
\author{Heiner Linke}
\author{Martin Leijnse}

\affiliation{Division of Solid State Physics and NanoLund, Lund University, Box 118,S-221 00 Lund, Sweden}

\maketitle

\section{Data on other devices}

\begin{figure}[H]
 \centering
 \includegraphics[width=0.90\linewidth]{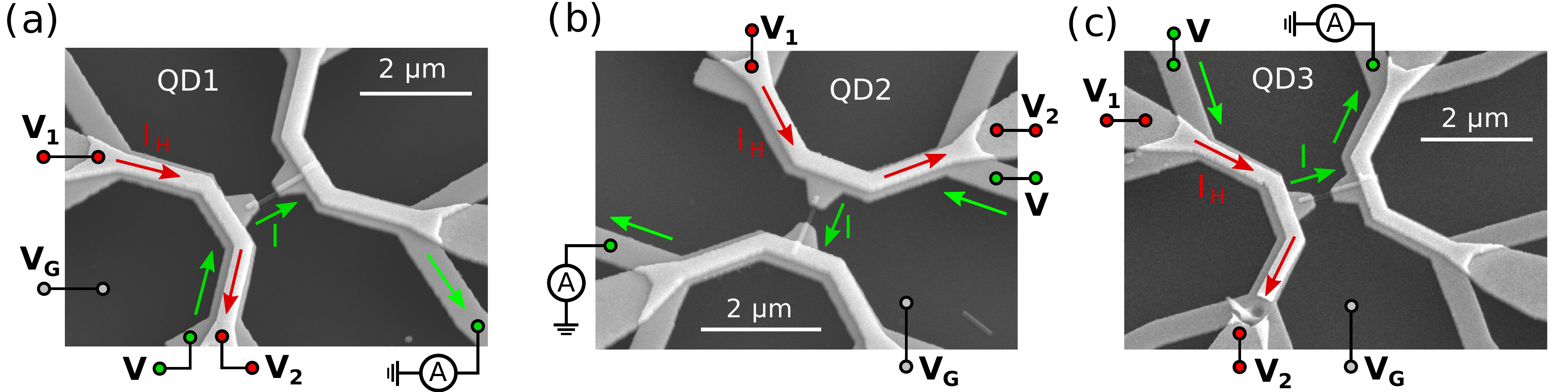}	
 \caption{Scanning electron micrograph (SEM) of devices QD1, QD2 and QD3. InAs/InP nanowires containing QDs are contacted by metallic leads on both sides of the QD. A bias voltage $V$ can be applied to the QDs via the leads. The path of the current $I$, resulting from $V>0$, is indicated by green arrows. The leads along with the nanowire are covered in $\mbox{HfO}_2$ to electrically insulate the overlaying thermal biasing circuit from the electrical circuit. A heater bias $V_H = V_2 - V_1$ is applied to one of the heater leads of every device resulting in a heating current $I_H$ that dissipates heat locally. The path and direction of $I_H$ resulting from $V_H>0$ is indicated by the red arrows. The damage to the heater visible on Device QD3 occurred during the process of unloading from the cryostat after the experiment was finalized.}
 \label{fig:Fig0}
\end{figure}

The main paper presents data and analysis on a certain resonance of device quantum dot 1 (labeled QD1a). In this section we provide additional data and the corresponding analyses on devices QD2 and QD3, as well as additional data on quantum dot 1 (labeled QD1b) measured over a different range of gate voltages. Figure S\ref{fig:Fig0} shows SEM images of the devices (QD1, QD2 and QD3). They all show clear signs of Kondo correlations in the charge stability diagrams (Fig.~S\ref{fig:Fig1}) when characterized at the base temperature of the cryostat $T_0<100$ mK. They also show similar behavior of the thermocurrent $I_{th}$ as a function of average temperature $T=(T_h+T_c)/2$  (Fig.~S\ref{fig:Fig3}), in particular a Kondo-related inversion of the direction of $I_{th}$ as a function of $T$. Complementary data on the behavior of $I_{th}$ in the presence of finite magnetic field $B$ is also given for QD1b and QD3. Unfortunately no data in the presence of magnetic field is available on QD2. All three devices are identical by design and are fabricated in the same process on the same sample chip. Figure S\ref{fig:Fig0} shows SEM images of the corresponding devices. 

\subsection{Accounting for series resistances in the measured differential conductance}

The current $I$ though the QD is measured by a current preamplifier with input impedance $R_I = 1 \mbox{ k}\Omega$. Both the contact and the heater leads are connected to the external electrical setup via DC lines with RC filters (total of $R_{RC} = 3.26 \mbox{ k}\Omega$ per line) designed to cut off frequencies above 300 Hz. In order to account for the serial resistances originating from the DC line filters and the input impedance of the current preamplifier when determining $g$, a distinction is made between $V$, which is the voltage across the QD, and $V_{ext}$, which is the voltage applied to the entire circuit (including the filter resistances and the input impedance of the preamplifier). Therefore, $g$ is calculated as $g = dI/(dV_{ext} - RdI)=dI/dV$, where $R = 2R_{RC} + R_I = 7.52\mbox{ k}\Omega$.

\begin{figure}[H]
 \centering
 \includegraphics[width=0.90\linewidth]{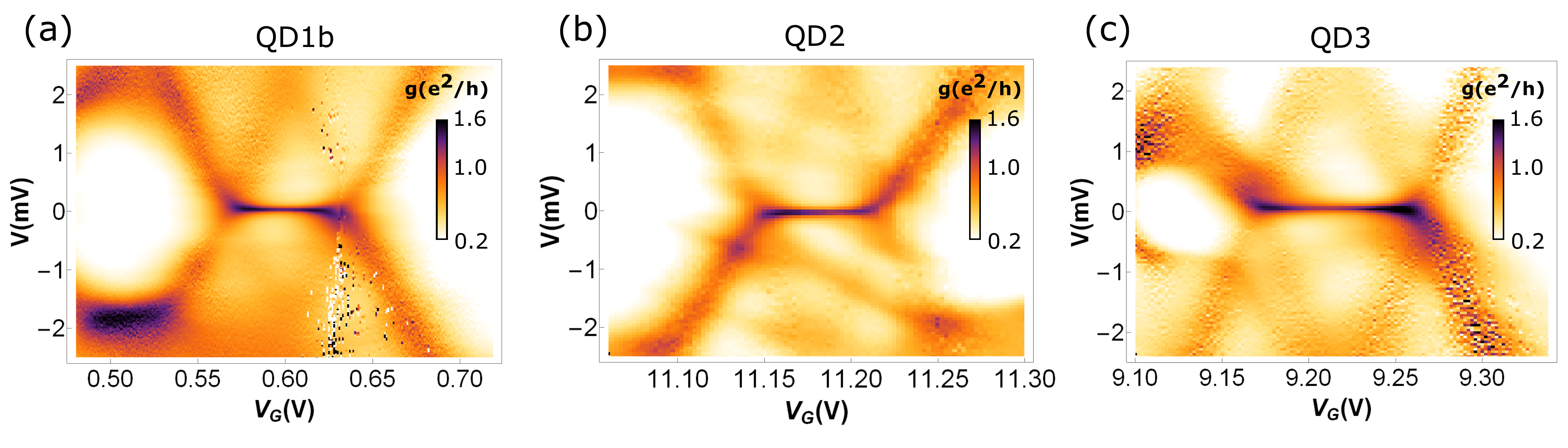}	
 \caption{$g$ as a function of $V_G$ and $V$ for Kondo resonances QD1b (a), QD2 (b) and QD3 (c). }
 \label{fig:Fig1}
\end{figure}

When taking the series resistance into account the resulting array of $g$ data for plotting charge stability diagrams is no longer rectangular in the coordinates of $V$ and $V_G$, therefore, polynomials %(of the order 3) 
are used to interpolate $g(V)$ in between the data points [concerns Fig.~3(c) in the main paper and Figs.~S\ref{fig:Fig1} and S\ref{fig:FigGf} here]. Figure S\ref{fig:Fig1} plots the charge stability diagrams measured on QD1b, QD2 and QD3 that show clear signs of Kondo correlations in the form of zero-bias conductance peaks (similar to resonance QD1a discussed in the main paper).

\subsection{Characterization of the Kondo temperature $T_K$ and the strength of the tunnel-coupling $\Gamma$.}

\begin{figure}[H]
 \centering
 \includegraphics[height=0.52\linewidth]{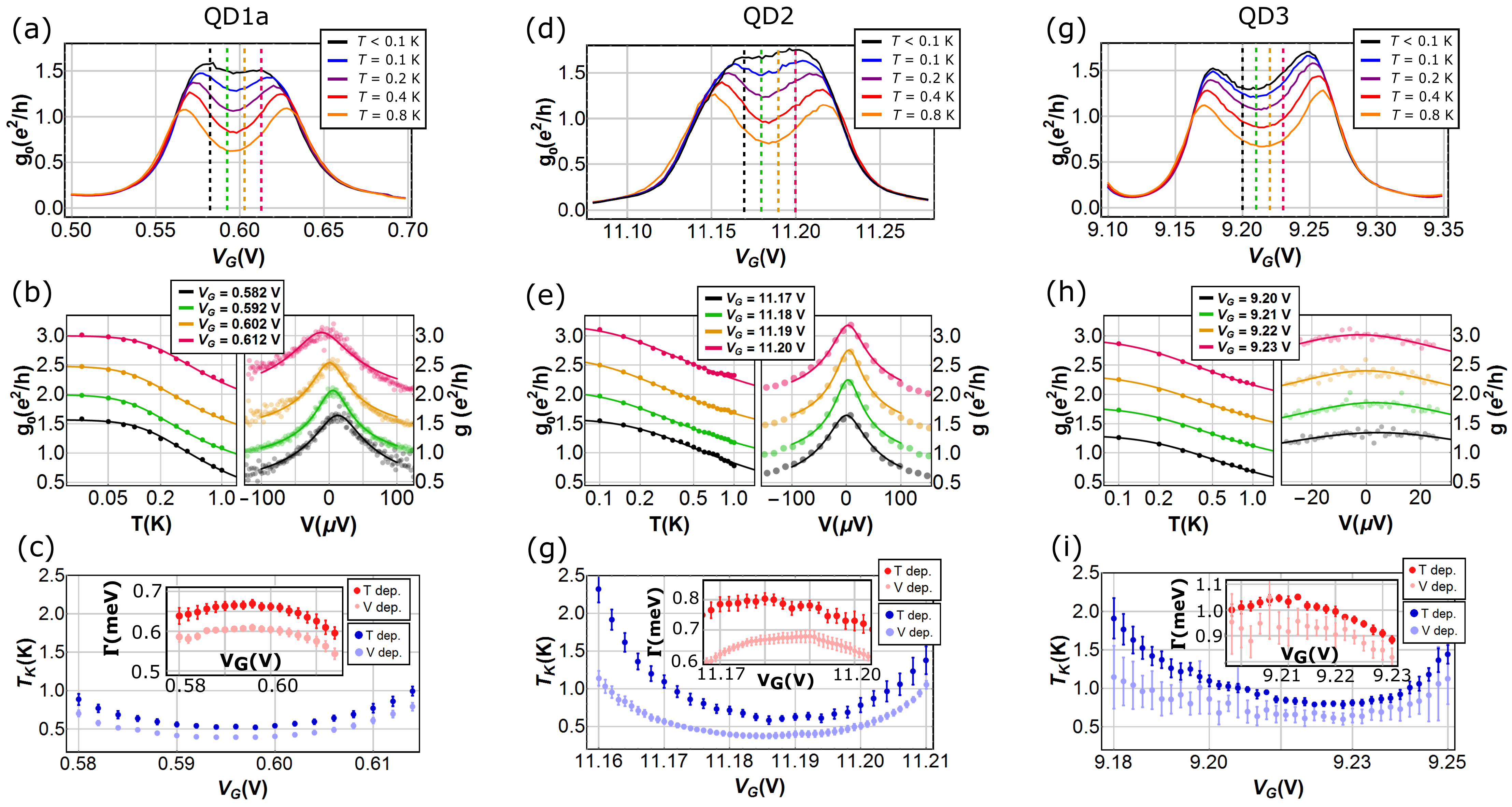}	
 \caption{Characterization of $T_K$ and $\Gamma$. (a)--(c) present data for QD1b, (d)--(f) present data for QD2, and (g)--(i) present data for QD3. (a), (d) and (g) show $g$ as a function of $V_G$ at different  $T$. (b), (e) and (h) show $g$ as a function of $T$ in left panels and as a function of $V$ in right panels, both for several different $V_G$ settings which are indicated in (a), (d) and (g) as dashed lines with the same color. The data points and curves are offset by 0.5 $e^2/h$. (c), (f) and (i) show $T_K$ results using the data from the $T$-dependence (dark blue points) and using the data from the $V$-dependence (light blue points). Error bars represent a $95 \%$ confidence interval for the $T_K$ value. Insets: results of numerical calculation of $\Gamma$ using the results for $T_K$ and Eq.~(\ref{eq:TK}) where $\varepsilon_0=  -e \alpha_G (V_G - V_{G0})$ with $\alpha_G$ being the gate lever arm. 
When calculating $\Gamma$, we used the following parameter values extracted from the charge stability diagrams: \textbf{QD1b}: $U = 2.2$ meV, $\alpha_G=0.0349$, $V_{G0}=0.564$ V; \textbf{QD2}: $U = 2.6$ meV, $\alpha_G=0.0382$, $V_{G0}=11.152$ V; \textbf{QD3}: $U = 3.0$ meV, $\alpha_G=0.0326$, $V_{G0}=9.169$ V.}
 \label{fig:Fig2}
\end{figure}

A characterization of the Kondo effect, equivalent to that presented in the main paper for QD1a, is also carried out for QD1b, QD2 and QD3. The characterization presented in Figs.~3 and~S\ref{fig:Fig2} is based on analyzing the behavior of $g$ as a function of $T$ and $V$. For the $T$ dependence we use the notation $g(V \approx 0) = g_0$ to distinguish it from $g(V)$. The quantity $g_0$ in Fig.~3 and Fig.~S\ref{fig:Fig2} is determined using linear fits to $I(V)$ within a range of $V\approx \pm 10\mbox{ }\mu\mbox{V}$, whereas $g(V)$ values are calculated as the differential increase in current $\Delta I$ in response to the voltage change $\Delta V$ between two sequential steps in $V$. We note that in the Fig.~2 the quantity $g_0$ is approximated by $I/V_{ext}$ where $V_{ext}= 25\mbox{ }\mu\mbox{V}$

The first row in Fig.~S\ref{fig:Fig2} shows how $g_0(V_G)$ evolves with increasing $T$. Within the Coulomb valleys $g_0$ demonstrates a gradual decrease with $T$ indicating the melting of the Kondo correlations. Examples of fits to $g_0(T)$ and $g(V)$ are shown in the second row of Fig.~S\ref{fig:Fig2}. The two dependencies are used to determine the value of the Kondo temperature $T_K$ by fitting it to empirical analytic expressions. For analyzing $g_0$, we use a standard expression
\begin{equation}\label{eq:g(T)}
g_0(T) = g_0(T=0) \left[ 1 + \left(2^{1/s}-1\right)\left(\frac{T}{T_K}\right)^2\right]^{-s},
\end{equation}
where $s=0.22$ for a spin-$1/2$ QD and $g_0$ and $T_K$ are used as free fit parameters \cite{Kretinin11,Pletyukhov2012}. For analyzing the $V$ dependence of $g$ we use the recently proposed expression \cite{Pletyukhov2012}
\begin{equation}\label{eq:g(V)}
g(V) = g_0(T=0) \left[ 1 + \left(\frac{2^{1/s}-1}{1-b+b\left(\frac{e(V-V_0)}{kT_K\prime}\right)^{s'}}\right)\left(\frac{e(V-V_0)}{kT_K\prime}\right)^2\right]^{-s},
\end{equation}
where $s=0.32$, $b=0.05$, $s'=1.26$ and the parameters $g_0(T=0)$, $V_0$ and $T_K$ are used as free fit parameters. Note that the qualitative shapes of $g_0(T)$ and $g(V)$, as defined by in Eqs.~(\ref{eq:g(T)}) and (\ref{eq:g(V)}), have a slightly different qualitative behavior. Also note that the two temperature constants, $T_K$ and $T_K\prime$, do not have the same meaning ($T_K \neq T_K\prime$). However, as pointed out in Ref.~\cite{Pletyukhov2012} using $T_K\prime \approx 1.8 T_K$ yields a definition of $T_K$ in Eq.~(\ref{eq:g(V)}) that is consistent with the definition of $T_K$ in Eq.~(\ref{eq:g(T)}), therefore we plot the results from $g(V)$ as $T_K \approx T_K\prime /1.8$.

The third row in Fig.~S\ref{fig:Fig2} shows the results for $T_K$ as a function of $V_G$. As expected, $T_K$ has a minimum in the center of the Coulomb valley. For a single-orbital model, $T_K$ is given by~\cite{Hewson93} [Eq.~(2) in the main paper]
\begin{equation}\label{eq:TK}
k_B T_K = \frac{1}{2}\sqrt{\Gamma U}\exp{\left({\frac{\pi \varepsilon_0(\varepsilon_0+U)}{\Gamma U}}\right)},
\end{equation}
which can be used to extract the corresponding tunnel-coupling strength $\Gamma$. The results of this calculation are shown as insets in the third row of the Fig.~S\ref{fig:Fig2}. The $V_G$ dependence of $\Gamma$ might be explained by an energy dependent density of states in the leads, which are effectively quasi one-dimensional pieces of nanowires.

\subsection{Thermoelectric characterization}

Thermoelectric characterization is realized by applying $\Delta T$ and measuring the thermocurrent $I_{th}$. In the linear response regime, and with no parasitic resistances in series with the QD, it is related to the thermovoltage $V_{th}$ via the relation $I_{th} = g_0 V_{th}$, therefore the detection of the sign inversion in $I_{th}$ also implies a sign inversion in $V_{th}$. Applying $\Delta T$ in our experimental devices also heats the colder reservoir. We use this effect to our advantage in order to study the $T=(T_h+T_c)/2$ dependence of $I_{th}$ as a function of $V_G$. All $I_{th}(V_G)$ traces (at each $T$) have been measured by sweeping $V_G$ in both directions a number of times (4 or 8) in order to average out current fluctuations related to the instability of the electrical bias $V$ across the QD with a typical magnitude of $\pm 1\mbox{ }\mu\mbox{V}$. Plots and the data analyses use the mean values of $I_{th}$ at every $V_G$.

\begin{figure}[H]
 \centering
 \includegraphics[height=0.48\linewidth]{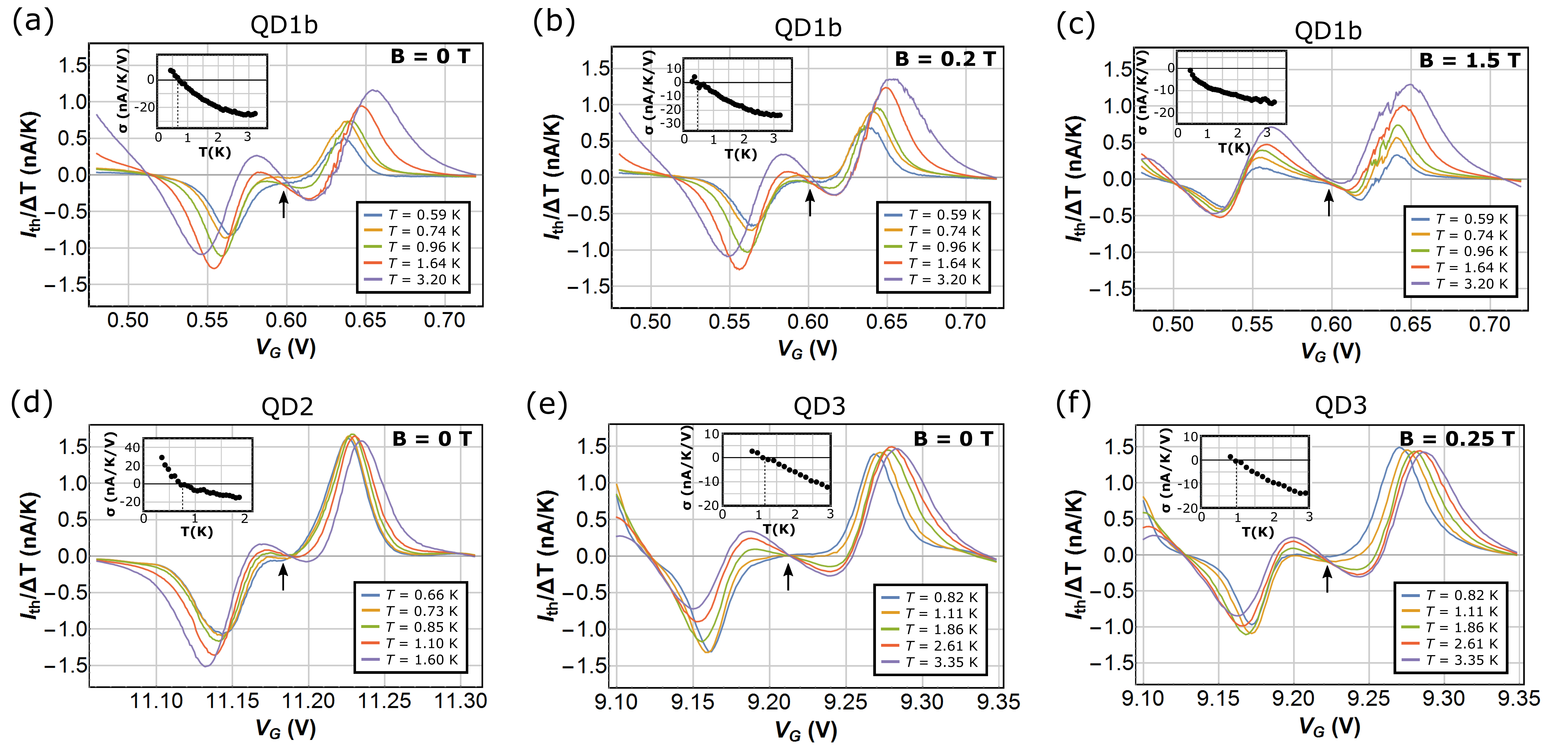}	
 \caption{Results of thermoelectric characterization. Results from QD1b are presented in (a) -- (c), results on QD2 are shown in (d), and results from QD3 are shown in (e) and (f). All plots present results on the measured $I_{th}$ normalized by $\Delta T = T_h - T_c$ for various $T = (T_h + T_c)/2$, as indicated in the figures. $B$ values are indicated in the right top corners of all subfigures (a)--(f). The insets present data for the corresponding thermocurrent slopes $\sigma = (dI_{th}/dV_G)/\Delta T$ that are determined at the $V_G$ values indicated by the black arrows. The temperature $T_1$ at which $\sigma$ crosses 0 is estimated from the graph and indicated by the vertical dashed lines.}
 \label{fig:Fig3}
\end{figure}

Figure S\ref{fig:Fig3} shows results of $I_{th}/\Delta T$ for QD1b, QD2 and QD3. In the absence of magnetic field ($B=0$) all three devices show a similar behavior to the one presented in the main paper, i.e., $I_{th}/\Delta T$ shows a sign inversion as a function of $T$ within the Coulomb valley where Kondo correlations can be observed at low $T$. We let $T_1$ denote the temperature at which the sign inversion takes place. Similarly as done in the main paper, we examine $\sigma = (dI_{th}/dV_G)/\Delta T$ as a function of $T$ to study the sign inversion.  To complement the data in the main paper, we also present data on QD1b and QD3 in presence of magnetic field $B$. For QD1b we present data at $B=0.2$~T and $B=1.5$~T [$ \sim 1.3 k_B T_K/(|g_Z| \mu_B)$ and $ \sim 10 k_B T_K/(|g_Z| \mu_B)$ respectively]. For QD3 we present data at $B=0.25$~T ($\sim 2.3 k_B T_K/(|g_Z| \mu_B)$). One can observe that for both (QD1b and QD3) the presence of a weak field [$B$ between 1 and 3 $k_B T_K/(|g_Z| \mu_B)$] does not reverse back the direction of $I_{th}$ which is consistent with observations in QD1a. However, the presence of the weak field seem to reduce $T_1$ instead of increasing it, as seen in presence of $ B\approx 3 k_B T_K/(|g_Z| \mu_B)$ for QD1a. In addition we show that under application of a strong field ($B=1.5$~T) to QD1b, the sign inversion is no longer observable and a behavior of $I_{th}/\Delta T$ characteristic to QDs without Kondo correlations is recovered, similarly as in the case of a strong field for QD1a [$B=2.0$~T, corresponding to $\sim 12 k_B T_K/(|g_Z| \mu_B)$] presented in the main paper.

\subsection{Measurements of g-factors}

\begin{figure}[H]
 \centering
 \includegraphics[height=0.25\linewidth]{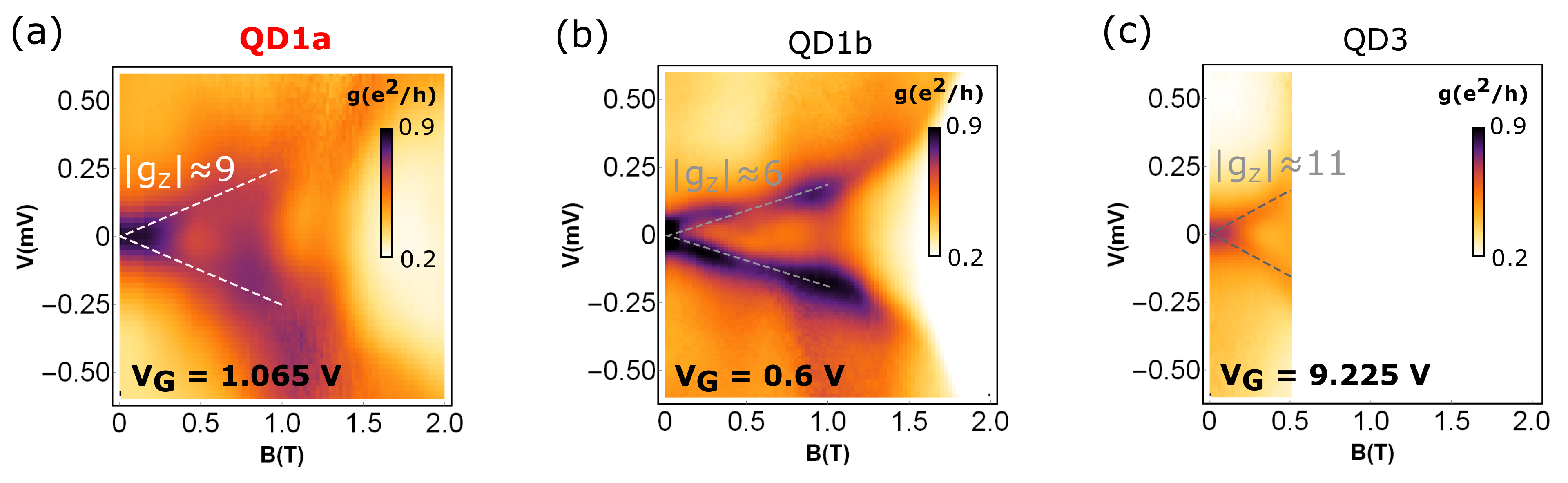}	
 \caption{Splitting of the Kondo peaks in magnetic field. All plots show differential conductance $g$ as a function of $V$ and $B$ (at $T_0< 100$~mK). (a) Data on QD1a, corresponding to the data the main paper. (b) Data on QD1b. (c) Data in QD3. The corresponding gate voltage settings for are indicated in figures. The dashed lines are rough estimates of the Zeeman splitting corresponding to the g-factor values, $|g_Z|$, that are also indicated in the figures.}
 \label{fig:FigGf}
\end{figure}

Figure~S\ref{fig:FigGf} shows the splitting of the Kondo peaks in magnetic field, based on which we estimate the g-factors. In all cases, we find values slightly below the bulk value $g_Z^\mathrm{bulk} \approx -14.9$ for InAs~\cite{Pidgeon67}. This is consistent with other studies showing a confinement-induced reduction and variation of g-factors~\cite{Winklerbook, Nilsson09}.

\section{Thermometry of QD3 in the weak coupling regime}

\begin{figure}[H]
 \centering
 \includegraphics[height=0.6\linewidth]{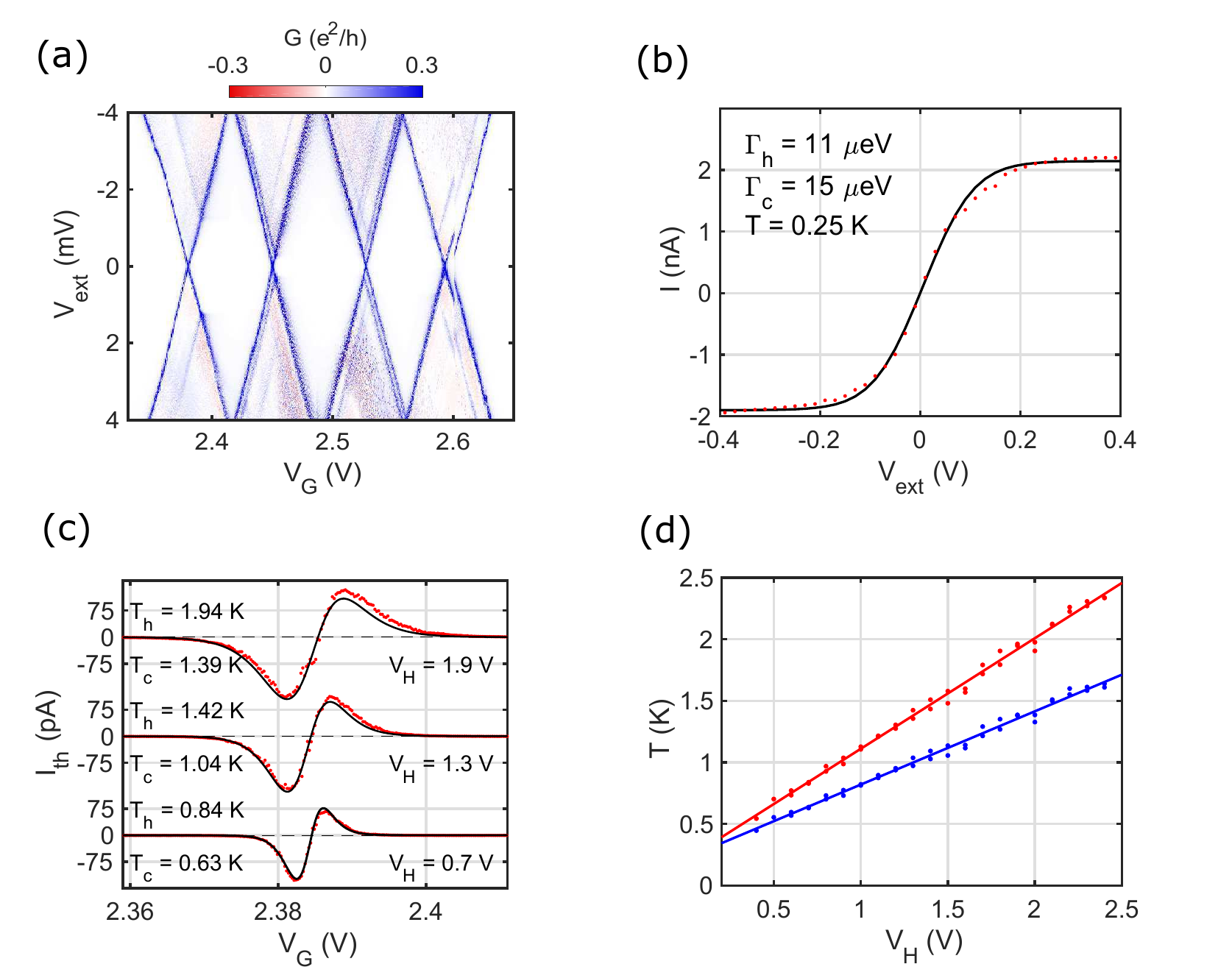}
 \caption{Temperature estimates using a weakly coupled resonance. (a) $G=dI/dV_{ext}$ as a function of $V_G$ and $V_{ext}$ show four possible resonances for temperature characterization. Only the resonance at $V_G = 2.385$ V was well reproduced by a single-level QD model with weak tunnel couplings. $U = 4.2$~meV and $\alpha_G = 0.061$ are determined from (a). (b) The tunnel rates are determined by fitting the RTD theory to the measured $I(V_{ext})$ at the resonance without thermal bias using $\Gamma_h$ and $\Gamma_c$ as free parameters. (c) Using the $\Gamma$s from (b) the temperatures of the electronic reservoirs are estimated by fitting the RTD theory to the measured $I_{th}(V_G)$ with $T_h$ (plotted in red) and $T_c$ (plotted in blue) as free parameters. The resulting temperatures and the applied $V_H$ are indicated in the figure. (d) The estimated temperatures for all $I_{th}(V_G)$ at this resonance show a linear increase as a function of $V_H$. Linear regression for the two temperatures yield $T_h=0.90V_H+0.21$ and $T_c=0.60V_H+0.22$. All measurements and calculations in the figure include a series resistance $R=7.52\ \text{k}\Omega$.}
 \label{fig:thermometry}
\end{figure}

To estimate the elevated electronic temperatures induced by the current $I_H$ though the heater lead we use the thermocurrent $I_{th}$ as a function of $V_G$ in the weak coupling regime ($\Gamma \ll k_B T$) where Kondo correlations are absent. The measured $I_{th}(V_G)$ is sensitive to both $T_c$ and $T_h$ and therefore can be compared to theoretical calculations to estimate the temperatures, in the same way as done in Ref.~\cite{Josefsson&Svilans17}. Such measurements were performed on all QDs, however, only QD3 displayed a resonance which exhibited the type of clear single-level behavior and weak enough coupling needed for the thermometry approach to be valid. Hence, this data is used to estimate $T_h$ and $T_c$ as a function of the heater bias $V_H$. We find an approximately linear dependence of temperatures on $V_H$ which is consistent with observations in previous studies~\cite{Josefsson&Svilans17} (see Fig.~S\ref{fig:thermometry}). Unfortunately, the thermometry data on QD3 does not cover the full range of $V_H$ used in measurements on other devices, and we therefore base our temperature estimates for large $V_H$ on a linear extrapolation of $T_h$ and $T_c$. Because all devices are identical by design we apply the thermometry result of QD3 also to devices QD1 and QD2. The estimated $\Delta T$ yields consistent results for $I_{th}/\Delta T$ on all devices which increases our confidence that this is a good approximation. 

Our theoretical approach~\cite{Leijnse08a, Gergs2018} for calculating the current through the weakly coupled QD is based on the real time diagrammatic (RTD) approach~\cite{Konig97}, in which one expands the Liouville-von Neumann equation in $\Gamma$ in order to calculate the stationary state reduced density matrix of the QD, as well as the stationary charge current. We keep all terms up to order $\Gamma^2$ in the expansion which, in addition to sequential tunneling, also accounts for co-tunneling, fluctuations and energy renormalization processes. The resistive load in series with the QD is included in the modeling by solving the self-consistent equation $I_{th}(V_G,V)=-V/R$ (see Supplementary Information of Ref.~\cite{Josefsson&Svilans17}).

We model the QD as a single-orbital Anderson model. This Hamiltonian can be broken into the respective Hamiltonians for the electronic reservoirs ($H_R$), the QD ($H_D$) and the tunnel couplings ($H_T$)
\begin{equation}
	H = H_R + H_D + H_T.
\end{equation} 
Here the reservoirs are assumed to be non-interacting 
\begin{equation}
	H_R = \sum_{r=h,c}\epsilon_{r,\sigma,k}c^\dagger_{r,\sigma,k} c_{r,\sigma,k}, 
\end{equation}
and they are also assumed to be in local equilibrium at all times such that they can be characterized by the Fermi-Dirac distribution $f_r(\epsilon)=[e^{(\epsilon-\mu_r)/kT_r}+1]^{-1}$. $c^\dagger_{r,\sigma,k}$ ($c_{r,\sigma,k}$) is the creation (annihilation) operator for an electron in reservoir $r$ with wave vector $k$ and spin $\sigma$, and $\epsilon_{r,\sigma,k}$ is the eigenenergy for the same electron. The QD is modeled as a single spin-degenerate energy level with on-site electron-electron interactions 
\begin{equation}
	H_D = \sum_{\sigma}\epsilon_\sigma n_{\sigma}+Un_{\uparrow}n_{\downarrow}, \ \ \ n_{\sigma}=d^\dagger_\sigma d_\sigma,
\end{equation}
with single particle energy $\epsilon_\sigma$ and interaction strength $U$. The field operators acting on the QD subspace are denoted by the letter $d$. Finally the tunneling Hamiltonian is given by
\begin{equation}
	H_T = \sum_{r = h,c} t_{r,\sigma,k}c^\dagger_{r,\sigma,k}d_\sigma + \ h.c.\ .
\end{equation}
The amplitude for an electron making a tunneling transition $t_{r,\sigma,k}$ is related to the tunneling rate $\Gamma$ as 
\begin{equation}
	\Gamma_{r,\sigma} = 2\pi \nu_r |t_{r,\sigma,k}|^2,	
\end{equation}
where $\nu_r$ is the density of states of reservoir $r$.

The analysis required to obtain the temperatures is a three-step process. First, basic QD parameters are extracted from current measurements without a thermal bias. From the stability diagram in Fig. S\ref{fig:thermometry}(a) we find $U=4.2 $ meV and $\alpha_G = 0.061$. Since the Coulomb diamonds in Fig.~S\ref{fig:thermometry}(a) are almost perfectly straight we consider symmetric voltage drop across both reservoirs, $\mu_{h,c}=\pm V/2$. Next we determine $\Gamma_h$ and $\Gamma_c$ by fitting the RTD theory to the measured current as a function of applied voltage as the QD is gated to the resonant condition $V_G = 2.385$~V, see Fig.~S\ref{fig:thermometry}(b). Determination of $\Gamma$s is possible since the saturated current values for large $V_{ext}$, and thus also large $V$, are independent of other parameters, like temperature or series resistance \cite{Bonet02}. In the final step we fit the RTD theory to the measured $I_{th}(V_G)$ using the previously extracted values for $U,\ \alpha_G,\ \Gamma_h$ and $\Gamma_c$, see Fig.~S\ref{fig:thermometry}(c). This is done for all traces of $I_{th}(V_G)$, two for each $V_{H}$ setting.

\bibliography{cite}
\bibliographystyle{apsrev}